\documentclass[11pt]{article}
\usepackage{amsmath,amssymb,color,graphics,epsfig,cite}
\usepackage{amsfonts}
\newcommand{\be}{\begin{equation}}
\newcommand{\ee}{\end{equation}}
\newcommand{\bea}{\setlength\arraycolsep{2pt} \begin{eqnarray}}
\newcommand{\eea}{\end{eqnarray}}
\newcommand{\nn}{\nonumber}

\def\fft#1#2{{\frac{#1}{#2}}}

\def\0{{\sst{(0)}}}
\def\1{{\sst{(1)}}}
\def\2{{\sst{(2)}}}
\def\3{{\sst{(3)}}}
\def\4{{\sst{(4)}}}
\def\5{{\sst{(5)}}}
\def\6{{\sst{(6)}}}
\def\7{{\sst{(7)}}}
\def\8{{\sst{(8)}}}
\def\sst#1{{\scriptscriptstyle #1}}

\begin{document}

\begin{center}
{\Large {\bf A Note on Kerr/CFT and Wald Entropy Discrepancy\\ in High Derivative Gravities}}

\vspace{20pt}

{\large Hai-Shan Liu, H. L\"u}

\vspace{10pt}

{\it  Center for Joint Quantum Studies and Department of Physics,\\
School of Science, Tianjin University, Tianjin 300350, China}

\vspace{40pt}

\underline{ABSTRACT}
\end{center}

We examine the Kerr/CFT correspondence in Einstein gravity extended with quadratic curvature invariants. We consider two explicit examples in four and five dimensions and compute the central charges of the asymptotic symmetry algebras of the near horizon geometries, using the improved version of the BBC formalism that encompasses the information of the Lagrangian. We find that the resulting Cardy entropy differs from the Wald entropy, caused by the Riemann-squared $R^{\mu\nu\rho\sigma}R_{\mu\nu\rho\sigma}$ term.

\vfill{\footnotesize  hsliu.zju@gmail.com \ \ \ mrhonglu@gmail.com}

%\vfill {\footnotesize mrhonglu@gmail.com}

%{\footnotesize \hoch{*}Corresponding author}

\thispagestyle{empty}
\pagebreak

\section{Introduction}

One of the most intriguing and equally puzzling aspects of classical black holes is that they can emit Hawking thermal radiation through a semiclassical process \cite{bhthermo1}. It implies that a black hole is a thermodynamical system of mass (energy) \cite{Arnowitt:1960es}, temperature \cite{bhthermo1} and entropy \cite{bhthermo2}.  The microscopic interpretation of the Bekenstein-Hawking entropy, which is geometrical and one quarter of the area of the horizon,  has long been a challenge. The first major breakthrough was achieved by Strominger and Vafa in \cite{Stvf}, where the entropy of some special five-dimensional extremal black holes in string theory was derived by counting the degeneracy of the corresponding superymmetric soliton states. Many works have since been proposed and one line of the directions focuses on finding an underlying two-dimensional conformal field theory (CFT) to compute the black hole entropy by the Cardy formula \cite{cdf}. Strominger \cite{mics} showed that the entropy of the three-dimensional BTZ black hole \cite{btz} could be indeed counted by the application of the Cardy formula of the dual two-dimensional CFT \cite{Brown:1986nw}. Later, a Kerr/CFT correspondence was proposed \cite{kcft} that the four dimensional extremal Kerr black hole in Einstein gravity can be identified with a two-dimensional chiral CFT, and the entropy of the extremal Kerr black hole can be reproduced by applying the Cardy formula to the dual CFT. The Kerr/CFT correspondence has been widely tested to be robust in Einstein gravity with general minimally or non-minimally coupled matter.

The generalization of the Kerr/CFT correspondence to Einstein gravity extended with higher-order curvature polynomial invariants is subtle. The black hole entropy is no longer geometrical, but arises as the Neother charge and can be calculated using the Wald entropy formula \cite{Wald:1993nt}.  In order to reproduce this entropy using the Kerr/CFT correspondence, a key ingredient is the central charge of the conformal symmetry algebra of the near-horizon geometry of the extremal rotating black hole. The central charge calculated from the Barnich-Brandt-Comp\'ere (BBC) formalism \cite{bbc1,bbc2,bbc3} will necessarily miss the contributions from the total derivative terms in the Lagrangian since the BBC formalism is based on the equations of motion. A new improved version that encompasses the Lagrangian information was later developed \cite{highg}, and it appeared to have resolved the discrepancy between the entropy calculated from the Kerr/CFT correspondence and the Wald entropy formula.

Recently an analytic perturbative solution of five-dimensional rotating black hole with two equal angular momenta was constructed in Einstein-Gauss-Bonnet gravity and thermodynamical properties were analysed \cite{Ma:2020xwi}.  In the extremal limit, the Wald entropy is $S=2\pi J + 12 \alpha \pi^{5/3} J^{1/3}$. Inspired by this result, we would like to revisit the Kerr/CFT correspondence in Einstein gravity extended with quadratic Riemann tensor invariants. We find that the discrepancy between the Cardy and Wald entropies persists.

The note is organized as follows. In section 2, we give a brief review of the Kerr/CFT correspondence in Einstein gravity. In section 3, we review the method of \cite{highg} for general extremal rotating black holes in four dimensions and show that the discrepancy is caused by the Riemann squared invariant, associated with the non-vanishing of the integration of a total derivative term of the latitude angle $\theta$.  In section 4, we apply the method of \cite{highg} for the rotating black holes with equal angular momenta in five dimensions.  In this case, the near horizon geometry is homogeneous and we show that the discrepancy is again caused by the Riemann-squared term.  We end the note with a conclusion in section 5. In the appendix, we apply the original BBC formalism to the $D=5$ example and illustrate its deviation from the Wald entropy.

\section{A brief review of Kerr/CFT}

We begin with a brief review of the Kerr/CFT correspondence \cite{kcft} in a simple example, namely the four dimensional extremal Kerr black hole in Einstein gravity. The Kerr metric \cite{kerr1} contains two integration constants, the mass $M$ and angular momentum $J$. The metric becomes extremal when $M^2=J$.  (See e.g.~\cite{kerr2} for a review.) Since the Kerr/CFT correspondence probes the near-horizon geometry only, we shall thus not present the whole Kerr metric, but its near horizon metric. In the notation of \cite{kcft}, it is
\be
d\bar s^2 = 2 J \Omega^2 \big( -(1+r^2) d\tau^2 + \fft{dr^2}{1+r^2}+ d \tilde \theta^2 + \Lambda^2 ( d\varphi + r d\tau )^2 \Big)\,,\label{kerrnh}
\ee
where
\be
\Omega^2 = \fft{1+\cos^2 \tilde \theta}{2} \,, \qquad \Lambda = \fft{2 \sin \tilde \theta}{1 + \cos ^2\tilde \theta } \,.
\ee
The horizon is some distorted 2-sphere associated with coordinates $(\theta,\varphi)$ and the Bekenstein-Hawking entropy is simply given by one quarter of the area of the horizon
\be
 S = 2 \pi J \,.
\ee
The near-horizon geometry is described by a cohomogeneity-one metric, depending on the latitude angle $\tilde \theta\in [0,\pi]$, with the level surfaces of $U(1)$ bundles of AdS$_2$. Note that the AdS$_2$ metric is written in the global coordinates, rather than the original ``planar'' coordinates, namely
\be
ds_2^2=\fft{dy^2 - dt^2}{y^2}\,,\label{2dplanar}
\ee
arising naturally from the decoupling limit of the extremal Kerr metric. It was shown in \cite{kcft} that this near-horizon geometry preserves an asymptotic symmetry under certain appropriate boundary conditions and its generators form the Virasoro algebra. Explicitly, the metric (\ref{kerrnh}) is invariant asymptotically under the transformation generated by vector fields
\be
\xi_n = - e^{- i n \varphi} ( \partial \varphi + i n r \partial r ) \,,
\ee
where $n$ is an integer. The commutation relation of the vector fields can be easily calculated, giving
\be
i [\xi_m, \xi_n] = (m-n) \xi_{m+n} \,.
\ee
In order to compute the central charge of the symmetry algebra, one needs to start with the infinitesimal difference of the associated conserved charge that can be obtained from integrating the 2-form potential $k_\xi$, {\it i.e.~}
\be
\delta Q_\xi = \fft{1}{8\pi} \int_{\partial \Sigma} k_\xi \,,
\ee
over the $r\rightarrow \infty$ boundary of the spatial slice. For Einstein gravity, the 2-form $k_\xi$ can be derived from the BBC formalism \cite{bbc1,bbc2,bbc3}, and it is given by %\cite{kcft}
\bea
k_\xi[h,g] =&& - \fft14 \epsilon_{\alpha\beta\mu\nu} \Big[ \xi^\nu \nabla^\mu h -\xi^\nu \nabla_\sigma h^{\mu\sigma} +\xi_\sigma \nabla^\nu h^{\mu\sigma} + \fft12 h \nabla^\nu \xi^\mu - h^{\nu\sigma} \nabla_\sigma \xi^\mu \nn\\
&& + \fft12 h^{\sigma\nu}( \nabla^\mu \xi_\sigma + \nabla_\sigma \xi^\mu )  \Big] dx^\alpha \wedge dx^\beta \,,\label{einsteinkxi}
\eea
where $h_{\mu\nu}$ is a linear perturbation of the background metric $g_{\mu\nu}$.

The Dirac bracket algebra of the asymptotic symmetry is
\be
\{Q_{\xi_m}, Q_{\xi_n}\}_{\rm DB} = Q_{[\xi_m,\xi_n]} + \fft{1}{8 \pi} \int_{\partial \Sigma} k_{\xi_m} [ {\cal L}_{\xi_n} g, g ] \,,\label{qqalgebra}
\ee
where the second term gives the central charge and ${\cal L}_\xi g$ is the Lie derivative, associated with the background metric $g_{\mu\nu}$, with respect to vector $\xi$. For the near-horizon extremal Kerr metric (\ref{kerrnh}), the last term in the above equation can be calculated directly and is given by
\be
\fft{1}{8 \pi} \int_{\partial \Sigma} k_{\xi_m} [ {\cal L}_{\xi_n} g, g ] = - i (m^3+ 2 m) J \delta_{m+n,0} \,.
\ee
The associated quantum algebra is thus the Virasoro algebra
\be
[ L_m , L_n ] = (m-n) L_{m+n} + J m (m^2-1) \delta_{m+n,0} \,,
\ee
with the central charge proportional to the angular momentum
\be
c = 12 J \,.
\ee
For general non-extremal Kerr black holes, there are two horizons $r_\pm$, each giving rise to $T_\pm$ temperatures and they form $T_L$ and $T_R$:
\be
\fft{1}{T_\pm} = \fft{1}{T_L} \pm \fft{1}{T_R}\,.
\ee
In the extremal limit, $T_\pm$ coalesce and vanish, in which case $T_R$ vanishes whilst $T_L$ remains a constant, and can be interpreted as the vacuum temperature of the near-horizon geometry of the extremal Kerr black hole, whose AdS$_2$ is given by (\ref{2dplanar}).  In the Kerr/CFT correspondence, we are concerned with the metric (\ref{kerrnh}) where the AdS$_2$ factor is in global coordinates. The required coordinate transformation implies that the vacuum temperature now becomes the dimensionless Frolov-Thorne temperature $T_{\rm FT}=2M T_L$, which for the near horizon extremal Kerr geometry (\ref{kerrnh}) is simply
\be
T_{\rm FT} = \fft{1}{2 \pi} \,.
\ee
One can thus obtain the microscopic entropy of the extreme Kerr black hole via the Cardy formula \cite{kcft}
\be
S = \fft13 \pi^2 c\, T_{\rm FT} = 2 \pi J\,,
\ee
which agrees precisely with the Bekenstein-Hawking entropy.

\section{Kerr/CFT for higher derivative gravity theories}

In this section, we discuss the generalization of the Kerr/CFT correspondence to include higher-order Riemann curvature polynomial invariants. In Einstein gravity, both the temperature and entropy are geometrical, with the former proportional to the inverse of the Euclidean time period, and the latter one quarter of the area of the horizon. When higher-order curvature terms are involved, the Hawking temperature remains geometrical by the argument of Quantum Statistical Relation \cite{Gibbons:1976ue}, but the entropy is no longer geometrical, calculated by the Wald entropy formula as a Noether charge. The Kerr/CFT correspondence is analogous in this regard. In order to obtain the microscopic entropy of certain two-dimensional conformal field theory by the Cardy formula, we need to know both the temperature and the central charge of the Virasoro algebra. The Florov-Thorne temperature is again purely geometrical and determined solely by the metric.  By contrast, the calculation of the central charge requires the detailed knowledge of the theory.

   Intriguingly, for Einstein gravity, the form field $k_\xi$ (\ref{einsteinkxi}) obtained
from the BBC formalism, which is crucial for finding the central charge, is identical to that for calculating the conserved energy or mass obtained by Abbot and Deser \cite{Abbott:1981ff}. The AD formula was generalized to include the quadratic curvature invariants \cite{Deser:2002jk}. However, the formula is not expected to be relevant to the Kerr/CFT correspondence since the form $k_\xi$ field depends manifestly only on $R^2$ and $R^{\mu\nu}R_{\mu\nu}$ invariants, but not the full Gauss-Bonnet invariant, which certainly contributes to the Wald entropy.  Issues also arise in the BBC formalism since it is based on the symmetry of the equations of motion rather than the Lagrangian. For example, the Gauss-Bonnet term is topological in $D=4$ and does not contribute to the equations of motion and hence it would give no contribution to the entropy in the Kerr/CFT description based on the BBC formalism, as was pointed out in \cite{gbdis}. This problem was later resolved using an improved formalism based on the Lagrangian rather than its equations of motion, and a concrete example of a general extremal four-dimensional rotating black hole was explicitly demonstrated \cite{highg}. In this section, we review this example in great detail and point out a subtlety in the calculation such that the discrepancy between the Kerr/CFT correspondence and the Wald entropy remains.

\subsection{General formalism}

Following \cite{highg}, we consider a general class of gravity theories in $D$ dimensions, with Lagrangian density $\sqrt{-g} L$, where $L$ is a function of the metric and Riemman tensor, together with some minimally coupled matter, namely
\be
L=L(g_{\mu\nu}, R_{\mu\nu\rho\sigma}, {\rm matter})\,.
\ee
(Note that we adopt the convention that the $1/(16\pi)$ factor is included in $L$.) For simplicity, we shall not consider terms involving covariant derivatives of the Riemann tensor. To proceed, we assume that the Lagrangian admits some extremal rotating black hole, whose near horizon geometry contains a $U(1)$ bundle over AdS$_2$.  We assume further that a vector field that preserves the asymptotic boundary conditions exists and still has the form
\be
\xi_n = - e^{- i n \varphi} ( \partial \varphi + i n r \partial r ) \,,\label{generalkilling}
\ee
where $\varphi$ is the $U(1)$ fibre coordinate of period $2\pi$. Improving upon the BBC formalism, a new method of calculating the $(D-2)$-form potential $k_\xi$ was obtained in \cite{highg}. Focusing on the structure of $n^3$ term that is relevant for extracting the central charge, one has
\bea
\int_{\partial \Sigma} k_{\xi_n} [ {\cal L}_{\xi_{-n}} g, g ]\big|_{n^3} =&& \int_{\partial\Sigma} \Big( - 2 \big[ X_{\alpha\beta} {\cal L}_{\xi_{n}} \nabla^{\alpha}\xi^\beta_{-n} +({\cal L}_{\xi_{n}} X)_{\alpha\beta} \nabla^{[\alpha}\xi_{-n}^{\beta]} + {\cal L}_{\xi_{n}} W_\alpha \xi^\alpha_{-n} \big] \, \cr
  && - E[{\cal L}_{\xi_{n}} g,{\cal L}_{\xi_{-n}} g; g] \Big)\Big|_{n^3} \,,\label{genkformula}
\eea
where $|_{n^3}$ denotes the $n^3$ terms and ${\cal L}_\xi$ is the Lie derivative with respect to vector $\xi$. The quantities $X_{\alpha\beta} \,, W^\alpha$ and $E$ are defined by
\bea
&&(X_{\alpha\beta})_{\lambda_3\dots\lambda_D} = - \epsilon_{\rho\sigma\lambda_3\dots\lambda_D} Z^{\rho\sigma}{}_{\alpha\beta} \,,\qquad (W^\alpha)_{\lambda_3\dots\lambda_D} =   2 (\nabla_\beta X^{\alpha\beta} )_{\lambda_3\dots\lambda_D} \,, \cr
&&E_{\lambda_3\dots\lambda_D} = \epsilon_{\alpha\beta\lambda_3\dots\lambda_D} \fft12 ( - \fft32 Z^{\alpha\beta\rho\sigma} ({\cal L }_{\xi_n} g )_\rho{}^{\mu} ({\cal L }_{\xi_m} g )_{\mu\sigma} + 2 Z^{\alpha\rho\sigma\mu} ({\cal L }_{\xi_n} g )_{\rho\sigma} ({\cal L }_{\xi_m} g )^\beta_{\mu}) \,,
\eea
with
\be
Z^{\mu\nu\rho\sigma} = \fft{\partial  L}{\partial R_{\mu\nu\rho\sigma}} \,.
\ee
Note that the manifest $1/(8\pi)$ factor in (\ref{qqalgebra}) was absorbed into the definition of $L$ in this more general description.  In fact the difference between the BBC and this new formalism is the $E$-term above \cite{highg}.

\subsection{$D=4$}

In this note, we shall present two explicit examples of $D=4$ and $D=5$ dimensions.  In this
section, we focus on $D=4$. The near horizon geometry of general four dimensional extremal rotating black holes is a compact foliation of an $U(1)$ bundle over AdS$_2$, taking the form \cite{highg}
\be
ds^2 = A(\theta)^2 ( -r^2 dt^2 + \fft{dr^2}{r^2} ) + d\theta^2 + B(\theta)^2 ( d\varphi + j r dt )^2 \,,\label{gend4nearh}
\ee
where $j$ is an arbitrary constant. Without loss of generality, we take that the longitudinal $\varphi$ has a period of $2 \pi$. $A(\theta)$ and $B(\theta)$ are determined by equations of motion, which involve not only some higher-order curvature terms, but also matter fields. For the horizon to describe a 2-sphere, the latitude coordinate $\theta$ must be compact.  In other words, $\theta\in [\theta_1, \theta_2]$, where $\theta_1, \theta_2$ are the two adjacent roots of $B$ with $|B'(\theta_1)|=|B'(\theta_2)|=1$. The function $A(\theta)$ must be finite and nonvanishing for $\theta\in [\theta_1, \theta_2]$. The vector field that preserves the asymptotic boundary conditions is indeed (\ref{generalkilling}). (For the purpose of calculating the central charge, we need to take $r\rightarrow \infty$, hence there is no difference between $r^2$ and $r^2+1$.)

As we mentioned in the previous section, the calculation of Cardy entropy of an extremal rotating black hole through the Kerr/CFT correspondence involves two ingredients, the Frolov-Thorne temperature and the central charge. The Frolov-Thorne temperature is directly determined by the metric and is given by
\be
T_{\rm FT} = \fft{1}{2 \pi j} \,.
\ee
The central charge not only depends on the metric, but also the part of the theory that involves the curvature tensors. Explicitly the central charge is proportional to coefficient of the $n^3$ term in the form field $k_\xi$. The crucial step in computing central charge is to construct the explicit $k_\xi$ for the relevant gravity theory.

We now follow the steps of \cite{highg} and evaluate (\ref{genkformula}) for the metric (\ref{gend4nearh}). The formula (\ref{genkformula}) has four structures and we evaluate them one by one. The first term gives
\be
-2 \int_\Sigma X_{\alpha\beta} {\cal L}_{\xi_n} \nabla^\alpha\xi^\beta_{-n} = i n^3 j \int_\Sigma Z_{\mu\nu\rho\sigma} \epsilon^{\mu\nu}\epsilon^{\rho\sigma} \text{vol}(\Sigma)\,,
\ee
with vol$(\Sigma) = B(\theta) d\theta d\varphi$. Then, the central charge from the first term can be read off, giving
\be
c_1 = -12 j \int_\Sigma Z_{\mu\nu\rho\sigma} \epsilon^{\mu\nu}\epsilon^{\rho\sigma} \text{vol}(\Sigma)\,.
\ee
According to the Cardy formula, the entropy from the first term is given by
\be
S_1 = \fft{\pi^2}{3} c_1 T_{\rm FT} = - 2 \pi \int_\Sigma Z_{\mu\nu\rho\sigma} \epsilon^{\mu\nu}\epsilon^{\rho\sigma} \text{vol}(\Sigma) \,,
\ee
which is exactly the same as that of the Wald entropy formula \cite{highg}.  Thus in order for the Kerr/CFT correspondence matches the Wald entropy, the remaining three terms must add up to zero.

The second term of (\ref{genkformula}) gives
\be
- 2 \int_\Sigma ({\cal L}_{\xi_n} X)_{\alpha \beta} \nabla^{[\alpha}\xi_{-n}^{\beta]}\Big|_{n^3} = -4 i n^3 \int_\Sigma \Big(  k B(\theta ) (Z_{\hat{t}\hat{r}\hat{t}\hat{r}} - Z_{\hat{t}\hat{\varphi}\hat{t}\hat{\varphi}} )   - 2 Z_{\hat{t}\hat{\varphi} \hat{r}\hat{\theta}} A(\theta) A'(\theta) \Big)  \,,
\ee
where a prime denotes a derivative with respect to $\theta$, and $Z_{\hat{\mu}\hat{\nu}\hat{\rho}\hat{\sigma}}$'s are the components of tensor $Z$ in the vielbein base
\be
e^{\hat{t}} = r A(\theta) dt \,,\quad  e^{\hat{r}} = \fft{A(\theta)}{r} dr\,,\quad
 e^{\hat{\theta}} = d \theta \,, \quad  e^{\hat{\varphi}} = B(\theta) ( d \varphi + j r dt) \,.
\ee
The third term is
\bea
- 2 \int_\Sigma ( {\cal L}_{\xi_n} W_\alpha ) \xi_{-n}^\alpha \Big|_{n^3} &=& - 2 \int d\theta d \varphi \Big( 4 i A^2 \partial _\theta Z_{\hat{t}\hat{\varphi}\hat{r}\hat{\theta}}+ 2 i k B ( Z_{\hat{t}\hat{r}\hat{t}\hat{r}} + Z_{\hat{r}\hat{\varphi}\hat{r}\hat{\varphi}} )\cr
  &&- 4 i AA' (Z_{\hat{t}\hat{r}}\hat{\theta}\hat{\varphi} - Z_{\hat{t}\hat{\varphi}\hat{r}\hat{\theta}}) + 4 i \fft{A^2 B'}{B} ( Z_{\hat{t}\hat{\theta}\hat{r}\hat{\varphi}} + Z_{\hat{t}\hat{\varphi}\hat{r}\hat{\theta}} )\Big) \,.
\eea
The $SL(2,\mathbb R) \times U(1)$ isometry and the $t$-$\varphi$ reflection symmetry of the metric imply \cite{highg}
\be
Z_{\hat r \hat \varphi \hat r \hat \varphi} =- Z_{\hat t \hat \varphi \hat t \hat \varphi}\,, \quad Z_{\hat t \hat r \hat \theta \hat \varphi} = - 2 Z_{\hat t \hat \varphi \hat r \hat \theta} \,, \quad Z_{\hat t\hat \theta \hat r \varphi } = - Z_{\hat t \hat \varphi \hat r \hat \theta} \,.
\ee
With the above properties and integration by part, the third term turns out to be
\bea
- 2 \int_\Sigma ( {\cal L}_{\xi_n} W_\alpha ) \xi_{-n}^\alpha \Big|_{n^3} = - 2 n^3 \int_\Sigma d\theta d\varphi \Big( && 2 i k B( Z_{\hat t \hat r \hat t\hat r} - Z_{\hat t\hat \varphi \hat t \hat \varphi} ) + 4 i A A' Z_{\hat t \hat \varphi \hat r \hat \theta} \cr
 && + \partial_\theta (4 i A^2 Z_{\hat t \hat \varphi \hat r \hat \theta} ) \Big) \,.
\eea
It is important to note that this is where we begin to differ from \cite{highg}. Here, we do not follow \cite{highg} to throw away the total derivative term. Adding the second and third terms simplifies the expression, giving
\be
- 2\int_\Sigma \Big[ ({\cal L}_{\xi_n} X)_{\alpha \beta} \nabla^{[\alpha}\xi_{-n}^{\beta]}+ ( {\cal L}_{\xi_n} W_\alpha ) \xi_{-n}^\alpha \Big] \Big|_{n^3}
= - 8 i n^3 \int_\Sigma d\theta d \varphi \Big[ k B  ( Z_{\hat t \hat r \hat t \hat r} - Z_{\hat t \hat \varphi \hat t \hat \varphi} ) + \partial_\theta (A^2 Z_{\hat t \hat \varphi \hat r \hat \theta} ) \Big]\,.
\ee
The fourth term, i.e.~the $E$-term, is
\be
\int_\Sigma E[{\cal L}_{\xi_n} g, {\cal L}_{\xi_{-n}}g; g]\Big|_{n^3} = 8 i  n^3 \int_\Sigma d\theta d\varphi \, k  B(\theta) ( Z_{\hat t \hat r \hat t \hat r } - Z_{\hat t \hat \varphi \hat t \hat \varphi} ) \,.
\ee
One thus sees that there are marvelous cancellations among the second, third and $E$ terms. In whole, these three terms together turn out to be simply a total derivative:
\bea
&&- 2\int_\Sigma \Big[ ({\cal L}_{\xi_n} X)_{\alpha \beta} \nabla^{[\alpha}\xi_{-n}^{\beta]}+ ( {\cal L}_{\xi_n} W_\alpha ) \xi_{-n}^\alpha   + E[{\cal L}_{\xi_n} g, {\cal L}_{\xi_{-n}}g; g] \Big] \Big|_{n^3}  \cr
&&= - 8 i n^3 \int_\Sigma d\theta d \varphi \Big[ \partial_\theta (A^2 Z_{\hat t \hat \varphi \hat r \hat \theta} ) \Big] = - 16 \pi i n^3 A^2 Z_{\hat t \hat \varphi \hat r \hat \theta} \Big|^{\theta_2}_{\theta_1}\,.
\eea
We thus obtain the corresponding contribution to the total entropy from these three terms through the Cardy formula:
\be
S_{\rm extra} =  \fft{32 \pi^2}{j}  A^2 Z_{\hat t \hat \varphi \hat r \hat \theta} \Big|^{\theta_2}_{\theta_1}\,.\label{sextra0}
\ee
Ref.~\cite{highg} took the view that this term vanished identically, in which case the Kerr/CFT correspondence would agree precisely with the Wald entropy. However, we believe that the total derivative term doesn't vanish in general and we illustrate this with an explicit example in the next subsection.

\subsection{An explicit example}

Here, we consider the Einstein-Gauss-Bonnet theory in four dimensions, together with some minimally coupled matter:
\be
L =\fft{1}{16 \pi}\Big( R  + \alpha ( R_{\mu\nu\rho\sigma} R^{\mu\nu\rho\sigma} - 4 R_{\mu\nu} R^{\mu\nu} + R^2 ) + {\rm matter} \Big)\,.
\ee
Since the Gauss-Bonnet term is topological in $D=4$, it will not affect the equations of motion. However, it can affect the Wald entropy and also the Kerr/CFT correspondence based on \cite{highg}. For the Gauss-Bonnet term, the contribution to Kerr/CFT entropy from the {\it total $\theta$-derivative term} is
\be
S_{\rm extra} = \fft{32 \pi^2}{k} A^2 Z_{\hat t \hat \phi \hat r \hat \theta}\Big|^{\theta_2}_{\theta_1} = \fft{4 \pi \alpha}{k} A^2   R_{\hat t \hat \phi \hat r \hat \theta} \Big|^{\theta_2}_{\theta_1} = - 2 \pi \alpha  ( \fft{d B}{d \theta} - \fft B A \fft{d A}{d \theta})  \Big|^{\theta_2}_{\theta_1} \,.
\ee
As was analysed earlier, for the horizon to be topologically a 2-sphere, we must have $B(\theta_1)=0=B(\theta_2)$.  This implies that
\be
S_{\rm extra}=  - 2 \pi \alpha   \fft{d B}{d \theta}\Big|^{\theta_2}_{\theta_1}\,.\label{d4totalder}
\ee
This must not be zero, since $\fft{d B}{d \theta}$ must be equal in magnitude but opposite in sign at the two adjacent roots $\theta_{1,2}$, such that the two-space of $(\theta,\varphi)$ is absent from any conic singularity.

Now, we consider an explicit example, by turning off all the matter fields, so that the vacuum solution is Ricci flat.  The near horizon geometry of the extremal Ricci-flat Kerr black hole was given in (\ref{kerrnh}). Applying for the Wald entropy formula, it is straightforward to obtain
\be
S_{\rm W} = 2 \pi  ( J + 2 \alpha) \,.
\ee
In other words, although the Gauss-Bonnet term has no effect on the solution, it can modify the Wald entropy.

We now examine how this topological term affect the Kerr/CFT correspondence. For the metric (\ref{kerrnh}), the explicit form of $A$ and $B$ are
\be
A(\theta) = \sqrt {J( 1+ \cos^2 \tilde \theta) } \,, \qquad B(\theta) = \fft{2 \sqrt J \sin \tilde \theta}{\sqrt { 1 + \cos^2 \tilde \theta }} \,,
\ee
and $j=1$.  The coordinates $\theta$ and $\tilde \theta$ are related by
\be
d \theta = \sqrt {J( 1+ \cos^2 \tilde \theta) } d \tilde \theta \,.
\ee
Therefore, we have
\be
\fft{dB}{d\theta} = \fft{1}{\sqrt {J( 1+ \cos^2 \tilde \theta) }}\fft{dB}{d\tilde \theta}\,, \qquad
\fft{dA}{d\theta} = \fft{1}{\sqrt {J( 1+ \cos^2 \tilde \theta) }}\fft{dA}{d\tilde \theta} \,.
\ee
This implies that
\be
S_{\rm extra} = -2\pi \alpha   ( \fft{d B}{d \theta} - \fft B A \fft{d A}{d \theta})  \Big|^{\theta_2}_{\theta_1} = -2\pi \alpha \Big(  \fft{2 \cos \tilde \theta}{1+\cos^2 \tilde \theta} - \fft{4 \sin^2 \tilde \theta \cos \tilde \theta}{(1+\cos^2 \tilde \theta)^2} \Big)\Big|^{\tilde \theta = \pi}_{\tilde \theta = 0} = 4 \pi \alpha \,.
\ee
Thus, we see that the microscopic entropy calculated from the Cardy formula based on the Kerr/CFT correspondence, with all contributions included,  is
\be
S_{\rm C} = 2 \pi (J + 4 \alpha) \,.
\ee
It is not equal to $S_{\rm W}$ and the discrepancy is due to the contribution to the entropy from the total $\theta$-derivative term (\ref{d4totalder}) that doesn't vanish:
\be
S_{\rm C} = S_{\rm W} + S_{\rm extra}\,.
\ee
It is worth pointing out that the extra term (\ref{sextra0}) involves the $Z$ tensor of all four directions $(\hat t, \hat \varphi, \hat r, \hat \theta)$. This implies that the covariant invariants involving only the Ricci scalar or Ricci tensor or at most a linear Riemann tensor will not contribute to $S_{\rm extra}$. Thus the quadratic terms $R^2$ and $R^{\mu\nu}R_{\mu\nu}$, which can be generated perturbatively by field redefinition $g_{\mu\nu} \rightarrow \gamma_1 R_{\mu\nu} + \gamma_2 R g_{\mu\nu}$, give no contribution to $S_{\rm extra}$, whilst the Riemann-squared term $R^{\mu\nu\rho\sigma} R_{\mu\nu\rho\sigma}$ will certainly do.

It can be argued that the entropy contribution from the Gauss-Bonnet term in four dimensions is trivial since it only adds a pure constant $\alpha$ of some numerical factor to the total entropy and it appears to be much ado about nothing to be fastidious on this issue. One thus needs to examine the Gauss-Bonnet effect in higher dimensions.  There is another reason to study this issue beyond four dimensions. The four-dimensional Kerr metric is cohomogeneity two, depending on both the radial $r$ and the latitude $\theta$ coordinates, with isometry of $U(1)\times \mathbb R$. When the metric becomes extremal, the near horizon geometry becomes cohomogeneity one, depending only on the latitude $\theta$ and the isometry is enhanced to $U(1)\times SL(2,\mathbb R)$. The explicit $\theta$ integration that leads to non-vanishing $S_{\rm extra}$ represents the inhomogeneous nature of the near-horizon geometry.  It is thus of great interest to investigate the situation when the near-horizon geometry is homogeneous and the angular integration in the central charge calculation yields only a universal volume factor of the horizon.  We shall carry out this in the next section.

\section{The Gauss-Bonnet contribution in five dimensions}

In this section, we study how the quadratic curvature invariants affect the Kerr/CFT in five spacetime dimensions.  The Kerr/CFT correspondence for Einstein gravity with or without a cosmological constant in diverse dimensions was studied in \cite{kcft1}, based on the general Kerr and Kerr-(A)dS black hole metrics \cite{Myers:1986un,Hawking:1998kw,Gibbons:2004uw,Gibbons:2004js}.  One motivation here is that the Gauss-Bonnet combination is no longer trivial and can affect the equations of motion.  The other motivation is that the near horizon geometry of the extremal Kerr black hole with the two equal angular momenta is homogeneous and we can thus avoid the analogous explicit $\theta$ integration in the $S_{\rm extra}$ term in four dimensions, discussed in the previous section.

We consider a more general quadratic extension, together with minimally-coupled matter
\be
L = \fft{1}{16 \pi}\Big( R + \alpha ( e_1 R^2  - 4 e_2 R_{\mu\nu} R^{\mu\nu} + e_3 R_{\mu\nu\rho\sigma}R^{\mu\nu\rho\sigma} ) + \hbox{matter} \Big) \,.
\ee
Here $(e_1,e_2,e_3)$ are three dimensionless constants, with the Gauss-Bonnet combination corresponding to $e_1=e_2=e_3=1$. In five dimensions, there can be two orthogonal rotations, and the metric ansatz can be very complicated for a generic theory.  When two angular momenta are equal, the metric becomes much simpler; it is cohomogeneity one, depending on the radial $r$ only, with the level surfaces being squashed 3-spheres, written as $U(1)$ bundles over $S^2$.  The most general ansatz is
\be
ds_5^2 = - \fft{h}{W} dt^2 + \fft{dr^2}{f} + \fft 14 W ( d \psi + \cos \theta d \phi + \omega dt  )^2 + \fft 14 r^2 ( d \theta^2 + \sin ^2 \theta d \phi ^2 ) \,,
\ee
where $h\,, f\,, W$ and $\omega $ are functions of $r$, and $\psi=2\varphi$ has a period of $4\pi$.  The extremal black hole is characterized by that $h$ and $f$ have the same double root at $r=r_+$:
\be
h =h_2 (r - r_+)^2 + {\cal O}(r-r_+)^3+ \cdots\,,\qquad  f = f_2 (r - r_+)^2 + {\cal O}(r-r_+)^3+ \cdots\,.
\ee
The near-horizon geometry can then be cast into
\bea
ds^2_5 = && \fft{1}{f_2} \Big(  -(1+r^2) dt^2 + \fft{dr^2}{1+r^2}  \Big) + \fft14{r_+^2} W(r_+) ( d \psi + \cos \theta d\phi + \sqrt{\fft{W(r_+)}{f_2 h_2}} \omega'(r_+) r dt )^2 \cr
 && + \fft 14 r_+^2 ( d \theta^2 + \sin ^2 \theta d \phi ^2 )  \,.\label{d5nearh}
\eea
The metric is homogeneous describing a constant $U(1)$ bundle over AdS$_2\times S^2$. The Frolov-Thorne temperature is
\be
T_{\rm FT} = \fft{\sqrt{f_2 h_2}}{ \pi \omega'(r_+) \sqrt{W(r_+)}} \,.
\ee
The entropy of the rotating black hole can be calculated by the Wald entropy formula and is given by
\bea
S_{\rm W} &=&  \fft12 \pi^2 r_+^3\sqrt {W(r_+)}  + \fft{\alpha \pi^2 r_+ \sqrt{W(r_+)}}{8 h_2}  \Big[ 16 e_1 h_2 (4- W(r_+))\cr
 &&- 16 (e_1 - 2 e_2 + e_3) f_2 h_2 r_+^2 + ( e_1 - 4 e_2 + 3 e_3) f_2 r_+^4 W(r_+)^2 \omega'(r_+)^2 \Big] \,.
\eea
Following the same method described in the previous section, we find that the entropy from the first term of the 3-form $k_\xi$ is the same as that of the Wald entropy, as was pointed out in general by \cite{highg}. The contributions to entropy from the rest three terms are
\bea
S_2+S_3 &=&  \pi ^2 \alpha  \fft{ r_+ \sqrt{ W(r_+)}}{h_2}  \Big[ 4 (e_3-e_2) f_2 h_2 r_+^2 + (e_2-e_3)f_2 r_+^4 W(r_+)^2 \omega '(r_+)^2 \cr
&& \qquad \qquad \qquad \qquad +(4 e_3-8 e_2) h_2 W(r_+) \Big]\,, \cr
S_E &=& -  \pi ^2 \alpha  \fft{r_+  \sqrt{ W(r_+)}}{h_2} \big[ 4 (e_3 - e_2) f_2 h_2 r_+^2 + (e_2-e_3)  f_2 r_+^4 W(r_+)^2 \omega '(r_+)^2 \cr
  && \qquad \qquad \qquad \qquad -8 e_2 h_2  W(r_+) \big] \,.\label{S23SE}
\eea
These three terms in whole have remarkable cancelations; however, they do not vanish:
\be
S_{\rm extra}=S_2 + S_3 +S_E = 4 e_3 \pi^2 \alpha r_+  W(r_+)^{\fft32} \,.
\ee
The total entropy calculated from the Cardy formula is thus
\be
S_{\rm C} = S_{\rm W} + 4 \pi^2 e_3 \alpha\, r_+  W(r_+)^{\fft32} \,.
\ee
As in the earlier four-dimensional example, the discrepancy is proportional to
the coefficient of $R^{\mu\nu\rho\sigma} R_{\mu\nu\rho\sigma}$.  Thus the Kerr/CFT correspondence works perfectly with the $R^2$ and $R_{\mu\nu} R^{\mu\nu}$ extensions, but not the $R^{\mu\nu\rho\sigma} R_{\mu\nu\rho\sigma}$ term. The result is consistent with the fact that the former can be generated by the metric field redefinition whilst the latter cannot be.
As a contrast, it follows from (\ref{S23SE}) that the original BBC formalism, for which $S_E$ is absent, will produce discrepancy even for the $R^{\mu\nu} R_{\mu\nu}$ term as well.
Note that the absence of the $e_1$ term in (\ref{S23SE}) can be understood that the $R^2$ term is equivalent to introducing a scalar to Einstein theory and hence we expect $S_{\rm C}=S_{\rm W}$ for the $R + \alpha R^2$ theory in both the BBC formalism and the improved version.

As a specific example, we may set $e_1 = e_2 = e_3 =1$, the theory becomes Einstein-Gauss-Bonnet gravity, and the corresponding Wald entropy has a much simpler form
\be
S^{\rm GB}_{\rm W} = \fft12 \pi^2 r_+^3 \sqrt {W(r_+)}  +2 \pi^2 \alpha r_+ \sqrt {W(r_+)} ( 4 -  W(r_+) ) \,.
\ee
While the contribution to entropy from the last three terms in the $k$-potential have also tremendous cancellations and (\ref{S23SE}) turns out to be
\be
S_2 +S_3 = - 4 \pi ^2 \alpha   r_+ W^{\fft 32}(r_+)\,, \qquad S_E = 8 \pi ^2 \alpha   r_+ W^{\fft 32}(r_+)\,.
\ee
It is worth pointing out that if one put erroneously an extra overall $1/2$ factor to the $E$-term by hand, then last three terms would have cancelled out completely, such that the Kerr/CFT correspondence would match the Wald entropy, as it was done in \cite{gbbmpv}. However, this ``miracle'' is coincidental for a specific example; it disappears for the general $(e_1,e_2,e_3)$ parameters.

Together with the contribution from the first term that gives the correct Wald entropy, the total entropy calculated from the Cardy formula is
\be
S^{\rm GB}_{\rm C} = \fft12 \pi^2 r_+^3 \sqrt {W(r_+)}  + 2 \pi^2 \alpha  r_+ \sqrt {W(r_+)} (4 +  W(r_+) )\,.
\ee
The discrepancy is
\be
S^{\rm GB}_{\rm C} - S^{\rm GB}_{\rm W} =  4 \pi^2 \alpha r_+ W(r_+)^{\fft32} \,.
\ee

To end this section, we would like to emphasize that the difference between the BBC formalism and its improved version differs by the contribution from the $E$-term.  In the appendix, we present an independent calculation based on the original BBC formalism to confirm our result. With these we conclude that the discrepancy has not been resolved.

\section{Conclusion}

In this note, we examined the Kerr/CFT correspondence in Einstein gravities with minimally coupled matter, extended with three quadratic curvature invariants, namely $R^2, R^{\mu\nu} R_{\mu\nu}$ and $R^{\mu\nu\rho\sigma} R_{\mu\nu\rho\sigma}$.  Two ingredients are involved in the counting of the microscopic entropy by using the Cardy formula.  One is the Florov-Thorne temperature, which is geometrical and can be derived from the metric of the near-horizon geometry of the extremal rotating black hole.  The other is the central charge of the asymptotic symmetric algebra of the near-horizon geometry.  We applied both the BBC formalism and its improved version that encompasses the information of the Lagrangian to compute the central charge so that we derived the Cardy entropy and compared it to the Wald entropy.

Explicitly, we considered two concrete examples of near-horizon geometry. One is in four dimensions and it is the most general cohomogeneity-1 metric of the compact foliation of the $U(1)$ bundle over AdS$_2$. The other is five dimensional and homogeneous, describing a constant $U(1)$ bundle over AdS$_2\times S^2$. Both are the near-horizon geometries of rotating black holes, with the latter having two equal angular momenta. We showed that there is a discrepancy between the Kerr/CFT correspondence and the Wald entropy. In particular, for the original BBC formalism, the discrepancy can arise from both $R^{\mu\nu} R_{\mu\nu}$ and $R^{\mu\nu\rho\sigma} R_{\mu\nu\rho\sigma}$ invariants.  (Adding the $R^2$ term is to effectively introduce a scalar to Einstein gravity and hence no deviation emerges.) If instead we adopt the improved version of the BBC formalism, some remarkable cancellations occur and the Card formula simplifies dramatically.  However, the discrepancy persists, but now caused by the Riemann-squared term $R^{\mu\nu\rho\sigma} R_{\mu\nu\rho\sigma}$ only. Our result (\ref{sextra0}) indicates a more general statement that the discrepancy arises when a curvature tensor polynomial term involves two or more Riemann tensors.  (Since the Riemann tensor can be expressed as Ricci tensors in three dimensions, such discrepancy we found in this note is not expected to be observed in three dimensions.)

Our results suggest that a new mechanism is needed to implement correctly the Kerr/CFT correspondence when higher-order curvature invariants are involved.  However, owing to the robustness of the BBC formalism and its improved version in other applications, the result might also suggest that the Kerr/CFT correspondence may fail to work in quantum gravity where higher-order curvature terms are quantum corrections that cannot be generated by the metric field redefinitions.

\section*{Acknowledgement}

We are grateful to the authors of \cite{highg} for correspondence and to Yi Pang, Junbao Wu and Jiaju Zhang for useful discussions. H.S.~Liu is supported in part by NSFC (National Natural Science Foundation of China) Grants No.~12075166 and No.~11675144.,  H.~L\"u.~is supported in part by NSFC Grants No. 11875200 and No. 11935009.

\appendix

\section{The Barnich-Brandt-Compere formalism}

In this appendix, we would like to calculate independently the form potential $k_{\xi}$ using the original BBC formalism. The BBC formalism and its improved version differ by the $E$-term in (\ref{genkformula}). This excise gives us a double check of the results in the main text. For a gravity theory with action $I$, the Einstein equation of motion is
\be
E_{\mu\nu}[g]  = \fft{\delta I}{\delta g^{\mu\nu}} \,.
\ee
For a small perturbation $g_{\mu\nu} = \bar g_{\mu\nu} + h_{\mu\nu}$ around the background metric $\bar g_{\mu\nu}$ that satisfies $E_{\mu\nu}[\bar g]=0$, we have the linearized tensor $E_{\mu\nu}^{(1)}$. With this tensor, we can define an asymptotic conservative current $S^\mu$ associated with an asymptotic
Killing vector $\xi^\mu$
\be
S_\xi^\mu[h; \bar g] = E^{(1)\mu\nu}[h;\bar g] \xi_\nu \,.
\ee
Then the form potential $k_\xi$ can be obtained by
\be
k^{\nu\mu}_\xi[h;\bar g] = \fft12 \phi^i \fft{\partial S^\mu_\xi}{\partial \phi^i_\nu} + ( \fft 23 \phi^i_\rho - \fft13 \phi^i \partial_\rho ) \fft{S^\mu_\xi}{\partial \phi^i_{\rho \nu}} + \dots - (\mu \longleftrightarrow \nu) \,,
\ee
with
\be
\phi^i = h_{\mu\nu} \,, \qquad \phi^i_\rho = h_{\mu\nu,\rho} \,, \qquad \phi^i_{\rho\sigma} = h_{\mu\nu,\rho\sigma} \,,
\ee
and so forth.  For theories involving higher-order curvature invariants, the calculation can be very involved and a mathematica code for this task was provide in \cite{bbc6}.  We use this code to calculate $k_\xi$ for the metric (\ref{d5nearh}) in Einstein-Gauss-Bonnet gravity and obtain the Cardy entropy
\be
S_{\rm C}^{\rm BBC} = \fft12 \pi^2 r_+^3 \sqrt {W(r_+)}  +2 \pi^2 \alpha r_+ \sqrt {W(r_+)} ( 4 - 3 W(r_+) )  \,.
\ee
We compare this with $S_{\rm W}$ and $S_{\rm C}$ discussed in the main text, and their relations are indeed
\be
S_{\rm C}^{\rm BBC} = S_{\rm W} -4 \pi ^2 \alpha   r_+ W^{\fft 32}(r_+)   = S_{\rm C} -8 \pi ^2 \alpha   r_+ W^{\fft 32}(r_+)  = S_{\rm C} - S_E \,.
\ee
Here $S_E$ is the entropy contribution from the $E$-term  in (\ref{genkformula}).
In fact, the simpler covariant expression of the form potential $k_\xi$ for the Gauss-Bonnet contribution was also obtained in \cite{Petrov:2009ns}, and we verify that it yields the same result above.   Note that this contribution vanishes in four dimensions, consistent with the fact that the BBC formalism is based on the equations of motion.

\end{document}